\documentclass[pra,showpacs,amsmath,amssymb,lengthcheck]{revtex4}

\usepackage{epsfig}

\begin{document}

\title{Trapped Two-Dimensional Fermi Gases with Population Imbalance}

\author{Bert Van Schaeybroeck$^{1,2}$, Achilleas Lazarides$^{3,4}$, Serghei Klimin$^{5}$
and Jacques Tempere$^{5}$} 

\affiliation{$^{1}$Koninklijk Meteorologisch Instituut (KMI), Ringlaan 3, B-1180 Brussels, Belgium}
\affiliation{$^{2}$Instituut voor Theoretische Fysica, Katholieke Universiteit Leuven, Celestijnenlaan 200 D, B-3001 Leuven, Belgium,}
\affiliation{$^{3}$Max Planck Institute for the Physics of Complex Systems, Noethnitzer Str.38, D-01187 Dresden, Germany}
\affiliation{$^{4}$Institute for Theoretical Physics, Utrecht University, Leuvenlaan 4, 3584 CE Utrecht, The Netherlands}
\affiliation{$^{5}$TFVS, Universiteit Antwerpen, Groenenborgerlaan 171, B2020 Antwerpen, Belgium} 

\date{\small\it \today}
\begin{abstract}
    We study population imbalanced Fermi mixtures under quasi-two-dimensional confinement at zero temperature. Using mean-field theory and the local-density approximation, we study the ground state configuration throughout the BEC-BCS crossover. We find the trapped system to be either fully normal or to consist of a superfluid core surrounded by a normal shell, which is itself either fully or partially polarized. Upon changing the trap imbalance, the trap configuration may undergo continuous transitions between the different ground states. Finally, we argue that thermal equilibration throughout the trap will be considerably slowed down at low temperatures when a superfluid phase is present.
\end{abstract}

\maketitle

\section{Introduction} The recent experimental progress on manipulating and cooling fermions has allowed the experimental realization and study of imbalanced fermion gases~\cite{partridge1,partridge2,zwierlein1,zwierlein2,zwierlein3,zwierlein4,zwierlein5}. An important experimental tool is the trapping of ultracold atoms in optical lattices using which various solid-state systems may be simulated, while the dimensionality of the system may be reduced to one or two dimensions~\cite{bloch}. However, perhaps the most extraordinary experimental possibility is the tunability of the interparticle interactions using Feshbach resonances~\cite{pethick,giorgini}. By adjusting the scattering length one is able to study the full BEC-BCS crossover in which weak interparticle attractions give rise to a Bardeen-Cooper-Schrieffer (BCS) type coupling while strong interactions lead to the formation of bound pairs which then undergo Bose-Einstein condensation (BEC).

Understanding the ground state properties of the \emph{three-dimensional} imbalanced three dimensional Fermi system is the subject of several recent theoretical works~\cite{sheehy,chevy,desilva2,gubbels,gubbels2,parish,haque,lobo,tempere2,diederix}. Zero-temperature Monte Carlo (MC) simulations~\cite{lobo}, renormalization group calculations~\cite{gubbels} and a model which incorporates beyond-mean-field and nonzero temperature fluctuations~\cite{tempere2} were all found to be in reasonable agreement with the MIT experiments~\cite{zwierlein1,zwierlein2,zwierlein3,zwierlein4,zwierlein5}. The Rice group's experiments~\cite{partridge1,partridge2}, on the other hand, appeared incompatible with the local-density approximation (LDA). Although surface effects were first believed to be the source of disagreement~\cite{haque,partridge2,desilva2,diederix}, the metastability of the experimental cloud system was eventually proven to provide the key to the understanding~\cite{hulet456,sommer}, as was anticipated in Refs.~\cite{vanschaeybroeck4,parish2}. Also recently, both theoretical and experimental studies of collective excitations with imbalanced fermion gases were performed~\cite{navon,lazarides,bruun,combescot,cooper}.

The physics of two-dimensional ultracold gases turns out to be significantly different from that of three dimensions. In practice, a quasi-2D system can be realized using optical lattices. This has already been done with \emph{balanced} Fermi mixtures~\cite{chin,bloch}. To date, theoretical studies of 2D imbalanced fermion gases have mainly focussed on homogeneous systems and were concerned with the effects of finite temperature~\cite{klimin44,desilva}, the Berezinskii-Kosterlitz-Thouless (BKT) transition~\cite{tempere3,lklim,tempere5,zhang2}, Fulde-Ferrel-Larkin-Ovchinnikov (FFLO) states~\cite{conduit,desilva}, interlayer tunneling~\cite{tempere4,desilva} and critical temperatures~\cite{petrov}.

In this work, we explore trapped, imbalanced 2D fermion gases at zero temperature using the mean-field theory for homogeneous phases of Ref.~\onlinecite{tempere1}. We present phase diagrams which depend only on an interaction parameter and the population imbalance. Three relevant trap configurations are identified: a fully normal configuration (FN) and two configurations consisting of a superfluid (SF) core surrounded by a fully polarized normal (FPN) or partially-polarized normal (PPN) shell, respectively. The transitions between these three configurations are of second order. Using two different approaches, we then investigate the collective excitations: we model the N phase both hydrodynamically and collisionlessly. The collective mode spectrum has frequencies below the trapping frequency. Finally, we find that the polarization of the N phase close to the interface has a strong effect on the thermal equilibration process: at low temperatures, energy transport through the N-SF interface is blocked.
\section{Trapped System}
\subsection{Trapping Potential}
Experimentally, a quasi-2D regime is established by combining tight confinement along the $z$-axis with weak harmonic trapping in the $x-y$-plane. The tight confinement is achieved using an one-dimensional optical lattice, which is modelled by a sinusoidal potential. The total trapping potential is:
\begin{align*}
    V({\bf r})=V_0 \sin^2\left(\frac{2\pi z}{\lambda}\right)+\frac{m\omega_0^2 r^2}{2},
\end{align*}
where $r^2=x^2+y^2$. If the energy cost of crossing the potential wells along the $z$-direction is much larger than the Fermi energy, the inter-well tunnelling rate is negligible and the particles are essentially confined in quasi-2D layers. In such a quasi-2D system a two-particle bound state always exists and has energy~\cite{randeria,petrov}:
\begin{align}
    \label{boundenergy}
    E_b=0.915 \hslash\omega_z \exp\left(\sqrt{2\pi}\ell_z/a\right),
\end{align}
where we have introduced $\omega_z^2=8\pi^2V_0/(m\lambda^2)$, $\ell_z^2=\hslash/m\omega_z$ and $a$ is the 3D s-wave scattering length. Within mean-field theory, the bound state energy $E_b$ fully characterizes the interparticle interactions and can be used to find qualitative results for the entire BEC-BCS transition. Therefore, as opposed to the 3D case, the crossover may be explored not only by tuning the scattering length $a$ but also by adjusting the trapping parameters $V_0$ and $\omega_z$.

\subsection{Equation of State}\label{eos} Let us now introduce the mean-field formalism which we will use later. We follow Ref.~\onlinecite{tempere1} in which homogeneous systems were thoroughly studied and analytic results were derived. This theory should be quantitatively correct in the BCS regime but also yields physically relevant results for the entire BEC-BCS crossover.

In the following, we consider two homogeneous phases: the normal (N) and the superfluid (SF) phase. The N phase is taken to consist of a non-interacting mixture of majority ($\uparrow$) and minority ($\downarrow$) particles. The N phase is said to be \emph{fully polarized} (FPN) when the density of the minority species is zero ($\rho_{\downarrow}=0$) and \emph{partially polarized} (PPN) otherwise. The interactions in the SF phase, on the other hand, are characterized by the bound state energy $E_b$ of Eq.~\eqref{boundenergy}. As was found both theoretically (Ref.~\onlinecite{tempere1}) and experimentally (Refs.~\cite{partridge1,partridge2,zwierlein1,zwierlein2,zwierlein3,zwierlein4,zwierlein5}), the two spin densities are equal in the SF phase. We use the local-density approximation in which, at each point in the trap, the results for the homogeneous phases are used.

At a point $\mathbf{r}$ in the trap, the pressure, $P$ and density, $\rho$, in the N ($i=\uparrow,\downarrow$) and SF phases are related by:
\begin{align}
    P_i(\mathbf{r})=\frac{\hslash^2\pi\rho_i^2(\mathbf{r})}{m}\quad\text{and}\quad P_{SF}(\mathbf{r})=\frac{\pi\hslash^2\rho_{SF}^2(\mathbf{r})}{2m}.
\end{align}
Also, the densities and chemical potentials $\mu$ obey the following relations:
\begin{align}
    \label{chemisch} \mu_i(\mathbf{r})=\frac{2\pi\hslash^2\rho_i(\mathbf{r})}{m}\text{ and } \mu_{SF}(\mathbf{r})=\frac{\pi\hslash^2\rho_{SF}(\mathbf{r})}{m}-\frac{E_b}{2}.
\end{align}

Within the LDA in the $x-y$-plane, $\mu(\mathbf{r})=\mu-m\omega^2_{0}r^2/2$ for all phases. We define the length scales $R_{\uparrow}$, $R_{\downarrow}$ and $R_{SF}$:
\begin{align}
    \label{lengten} \rho_i(\mathbf{r})=\frac{R_i^2-r^2}{4\pi \ell_{\omega}^4}\quad\text{and}\quad \rho_{SF}(\mathbf{r})=\frac{R^2_{SF}-r^2}{2\pi\ell_{\omega}^4},
\end{align}
with $i=\uparrow,\downarrow$ and $\ell_{\omega}=\sqrt{\hslash/m\omega_0}$. Finally, since the SF state is balanced, it consists of an equal $\uparrow$ and $\downarrow$ density at each point of the trap.

The N and the SF phases are separated by an interface at radial position $\zeta$ where two boundary conditions apply. First, mechanical equilibrium at the interface demands that the difference of the pressures at the interface must be compensated for by a surface-tension term~\cite{lazarides}. This is expressed by Laplace's formula~\footnote{Note that the pressure in this work has dimensions energy per unit surface. The surface tension has dimensions energy per unit length and, strictly speaking, should therefore be called line tension.}:
\begin{align}
    \label{laplace} \left(P_{SF}-P_{\uparrow}-P_{\downarrow}\right)_{\zeta}= \left(\frac{\sigma}{\zeta}\right)_\zeta,
\end{align}
where $\sigma$ is the surface tension and $(\cdot)_{\zeta}$ means that the quantity is evaluated at the interface. The second boundary condition pertains to the fact that pairs of opposite-spin particles may cross the interface~\cite{vanschaeybroeck4}. Therefore, at equilibrium, the energy to \emph{add} such a pair of particles in the N phase must be equal to the energy to \emph{break} a pair in the SF phase, or:
\begin{align}
    \label{chemicalequilibrium} 2\left(\mu_{SF}\right)_{\zeta}=\left(\mu_{\uparrow}+\mu_{\downarrow}\right)_{\zeta}.
\end{align}
In case the interface is between the SF and the FPN phase, no $\downarrow$ particles are present in the N phase. Therefore, in Eq.~\eqref{laplace} the pressure $P_{\downarrow}$ can be put to zero. We will later see that in that case, the interface becomes impermeable, that is, no particles can cross the interface, which is as it should be.

\subsection{Trap Characteristics} 
Using the \emph{local} properties mentioned above we can extract \emph{global} properties of the trap. We take the trap to contain $N=N_{\uparrow}+N_{\downarrow}$ particles in total. The particle numbers of the different spin species $N_{\uparrow}$ and $N_{\downarrow}$ are given by
\begin{align}
    \label{polarization} Q=(N_{\uparrow}-N_{\downarrow})/(N_{\uparrow}+N_{\downarrow}),
\end{align}
where $Q$ is the population imbalance. The particle densities of Eq.~\eqref{lengten} can be integrated to yield the particle numbers in the N shell, $N_{Ni}$, and in the SF core, $N_{SF}$:
\begin{subequations}
    \begin{align}
        N_{Ni}&=\frac{(R_i^2-\zeta^2)^2}{8\ell_{\omega}^4},\label{particlenormal}\\
        N_{SF}&=\frac{\zeta^2(R_{SF}^2-\zeta^2/2)}{2\ell_{\omega}^4},
    \end{align}
\end{subequations}
with $i=\uparrow,\downarrow$. Also, since the SF phase is unpolarized, the total particle number in each spin state is given by
\begin{align}
    N_{i}=N_{Ni}+N_{SF}/2.
\end{align}

We have now established all the ingredients necessary to construct all possible equilibrium states. This is done by solving the equations above for the variables $\zeta, R_{SF}$, $R_{\uparrow}$ and $R_{\downarrow}$. Three trap configurations turn out to be of relevance. First, a trap consisting of a SF core phase surrounded by a \emph{partially}-polarized normal (PPN) shell, which we denote by PPN+SF. A second configuration, denoted FPN+SF, consists of a SF core phase surrounded by a \emph{fully}-polarized normal shell~\cite{chevy}. Finally, the trap fully normal configuration is simply denoted by FN.

\begin{figure}
    \epsfig{figure=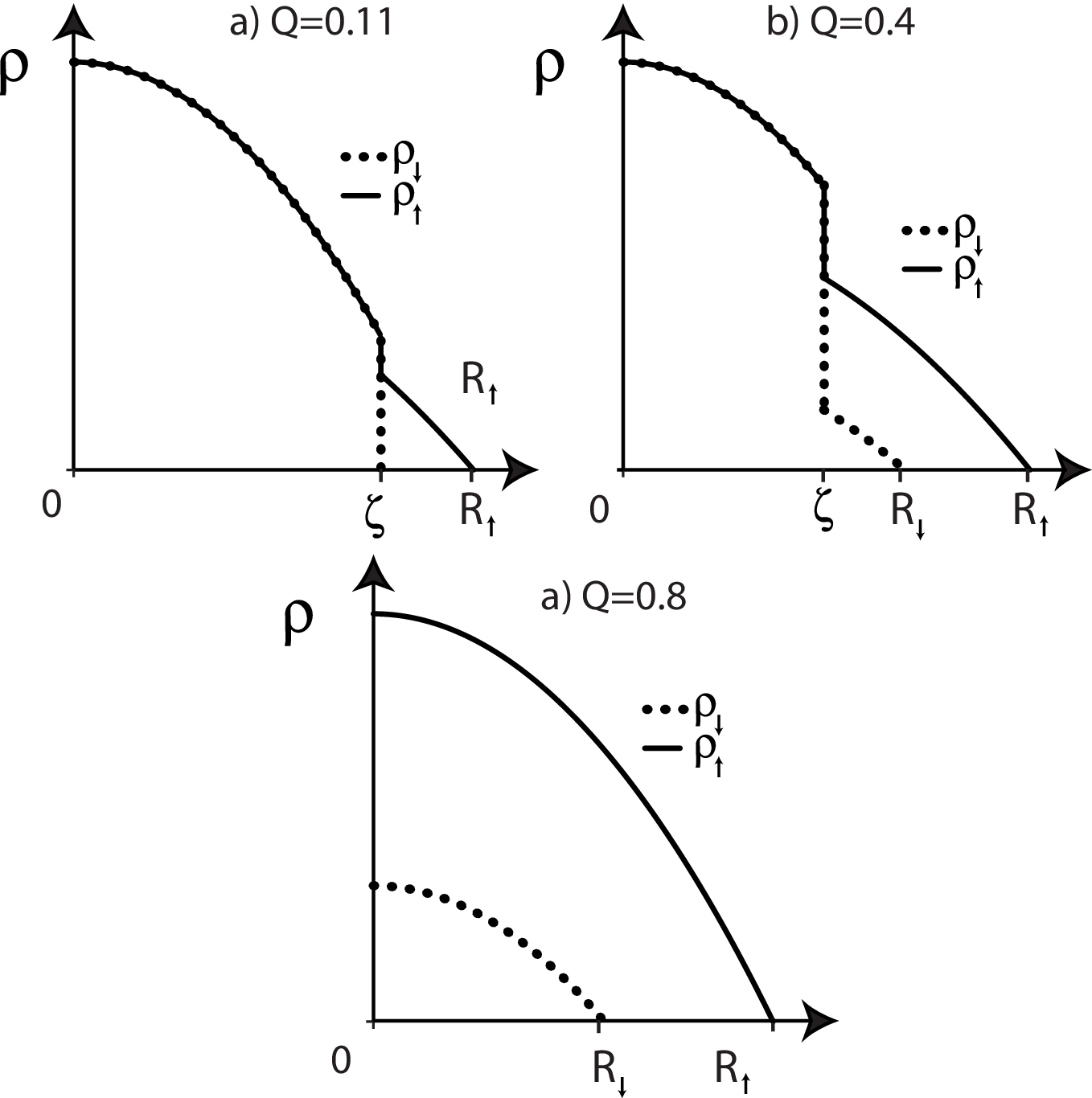,angle=0, width=240pt} 
\caption{The spin densities $\rho_{\uparrow}$ and $\rho_{\downarrow}$ of the three relevant trap configurations as a function of the radial coordinate $r$. The density of the majority species $\rho_{\uparrow}$ is drawn using a full line, while the minority species $\rho_{\downarrow}$ is shown with a dotted line. We took $E_b/E_F=0.2$ and varied the population imbalance $Q$. $\zeta$ is the position of the interface, while $R_{\uparrow}$ and $R_{\downarrow}$ are defined in Eq.~\eqref{lengten}. a) At low population imbalance (here $Q=0.11$) we find a configuration consisting of a SF core surrounded by a fully polarized normal shell (FPN+SF). b) At intermediate population imbalance (here $Q=0.4$) we find the trap to consist of a SF core surrounded by a partially-polarized normal shell (PPN+SF). c) At sufficiently high polarizations (here $Q=0.8$), the trap is fully normal (FN). \label{fig1}}
\end{figure}

In Fig.~\ref{fig1} the spin densities of these three trap configurations are shown. The minority spin density, $\rho_{\downarrow}$ vanishes in the FPN phase and $\rho_{\uparrow}=\rho_{\downarrow}$ in the SF phase.

\subsection{Trap Characteristics} 
We continue by giving some analytic results for zero surface tension, $\sigma=0$.

We begin with the PPN+SF configuration which, at the interface, is characterized by a partially-polarized normal phase: in terms of energy this means that $h<\left(\mu_{SF}\right)_\zeta$ where $h=(\mu_{\uparrow}-\mu_{\downarrow})/2$ is a constant throughout the trap. The coexistence condition Eq.~\eqref{laplace} and the particle number equation~\eqref{particlenormal} reduce to~\cite{tempere1}:
\begin{subequations}
    \begin{align}
        \label{gieter} h^2&=\left(\mu_{SF} E_b+\frac{E_b^2}{4}\right)_\zeta,\\
        N_{N\uparrow}-N_{N\downarrow}&=2\left(\frac{\mu_{SF} h}{(\hslash\omega_0)^2}\right)_\zeta.
    \end{align}
\end{subequations}
Combining these we find that $R_{SF}^2=2\ell_{\omega}^2\sqrt{N}$ and the interface position $\zeta$ can be obtained as a solution of the following equation:
\begin{align}
    \label{conditioning} 2Q^2E_F^4=&E_b\left(2E_F-E_b-m\omega_0^2\zeta^2\right)^2\nonumber\\
    &\times(2E_F-E_b/2-m\omega_0^2\zeta^2),
\end{align}
where we have defined the Fermi energy~\footnote{In fact, Eq.~\eqref{Fermi} is equal to the Fermi energies at the trap center of a FN configuration of a balanced fermion mixture of particle number $N$.}:
\begin{align}
    \label{Fermi} E_F=\hslash\omega_0 N^{1/2}.
\end{align}
Note that all parameters of Eq.~\eqref{conditioning} are \emph{global} trap characteristics.

In contrast to the PPN+SF configuration, exact analytic expressions may be obtained for the FPN+SF configuration. Full population imbalance at the N-SF interface implies that, at the interface, $h>\left(\mu_{SF}\right)_\zeta$. Therefore, Eqs.~\eqref{laplace} and~\eqref{particlenormal} reduce to~\cite{tempere1}:
\begin{subequations}
    \begin{align}
        \label{deurmat} h&=\left(\mu_{SF} (\sqrt{2}-1)+\frac{E_b}{\sqrt{2}}\right)_\zeta,\\
        2N_{N\uparrow}&=\left(\frac{\mu_{SF}+h}{\hslash\omega_0}\right)^2_\zeta.
    \end{align}
\end{subequations}
These equations result in the following radii:
\begin{subequations}
    \label{radiiPPN}
    \begin{align}
        R_{SF}^2&=2\ell_{\omega}^2\sqrt{N},\\
        R_{\uparrow}^2&=2\ell_{\omega}^2\sqrt{N}\left(1+\sqrt{Q}(\sqrt{2}-1)\right),\\
        \zeta^2&=2\ell_{\omega}^2\sqrt{N}\left(1-\sqrt{Q}\right),
    \end{align}
\end{subequations}
and clearly $\zeta\leq R_{SF}\leq R_{\uparrow}$ as it must. Note that $R_{SF}$ is independent of $Q$ and, remarkably, all lengths are independent of $E_b$. In the limit of a balanced trap ($Q=0$), all radii equal $\sqrt{2\sqrt{N}}\ell_{\omega}$; when, on the other hand, the trap is fully imbalanced the interface position vanishes while $R_{\uparrow}$ becomes equal to the Thomas-Fermi length $R_{TF}=\ell_{\omega}(8N)^{1/4}$.

\section{Phase Diagrams}

Having established the possible trap configurations we present now phase diagrams of trapped imbalanced 2D fermion gases at zero temperature. After introducing the concept of total trap energy in Sect.~\ref{energysection}, we discuss the phase diagram without (Sect.~\ref{noST}) and with surface tension.

\subsection{Total Trap Energy}\label{energysection} 
At zero temperature, whether or not a certain trap configuration is the ground state is determined by its total energy. Therefore, one must compare the energies of all three trap configurations at fixed total particle number $N$ and population imbalance $Q$.

The total energy is related to the grand-canonical thermodynamic potential
$\Omega_{tot}$ by the equation
\begin{equation}
E_{tot}=\Omega_{tot}+h\left(  N_{\uparrow}-N_{\downarrow}\right)  +\mu_{0}N,
\label{Etot}%
\end{equation}
where $\mu_{0}$ is the chemical potential of the whole trapped system. For the
PPN+SF and FPN+SF configurations, $\mu_{0}=\left.  \mu_{SF}\left(  r\right)
\right\vert _{r=0}$ whereas for the FN configuration, $\mu_{0}=\left.  \frac{1}%
{2}\left[  \mu_{\uparrow}\left(  r\right)  +\mu_{\downarrow}\left(  r\right)
\right]  \right\vert _{r=0}$.

By straightforward integration, we found the grand-canonical thermodynamic
potential as a sum of contributions of the normal shells $E_{Ni}$
($i=\uparrow,\downarrow$), the SF core phase $E_{SF}$ and the surface tension
$E_{\sigma}$ with:
\begin{subequations}
\label{Omega}%
\begin{align}
\Omega_{tot}  &  =\Omega_{SF}+\Omega_{\uparrow}+\Omega_{\downarrow
},\label{WTOT}\\
\Omega_{j}  &  =-\frac{\hslash\omega_{0}}{48l_{\omega}^{6}}\left(  R_{j}%
^{2}-\zeta^{2}\right)  ^{3},\label{W1}\\
\Omega_{SF}  &  =-\frac{\hslash\omega_{0}}{24l_{\omega}^{6}}\zeta^{2}\left(
\zeta^{4}-3\zeta^{2}R_{S}^{2}+3R_{S}^{4}\right)  ,\label{W2}\\
\Omega_{\sigma}  &  =2\pi\zeta\sigma. \label{W3}%
\end{align}
By addition of these contributions to Eq.~\eqref{Etot}, it is possible to
calculate the total energy of the three possible trap states. For instance,
the energy of the FN configuration can be obtained by setting $\zeta=0$ which
yields:
\end{subequations}
\[
E_{tot}=\frac{E_{F}}{3}\left(  \left(  1+Q\right)  ^{3/2}+\left(  1-Q\right)
^{3/2}\right)  .
\]
For the PPN+SF configuration with a zero surface tension, the total trap
energy is obtained using Eq.~\ref{Omega} with the
interface position $\zeta$ found by solving Eq.~\eqref{conditioning}.

Lastly, $E_{N\downarrow}$ must be set to zero when calculating the total trap
energy of the FPN+SF state. We find in case of a zero surface tension:
\[
E_{tot}=\frac{2E_{F}}{3}\left(  1+Q^{3/2}\left(  \sqrt{2}-1\right)  \right)
+\frac{E_{b}}{2}\left(  Q-1\right)  .
\]
These expressions for the energy allow us now to draw the phase diagram in
case of a zero surface tension.

\subsection{Phase Diagram with Zero Surface Tension}\label{noST} 
\begin{figure}
    \epsfig{figure=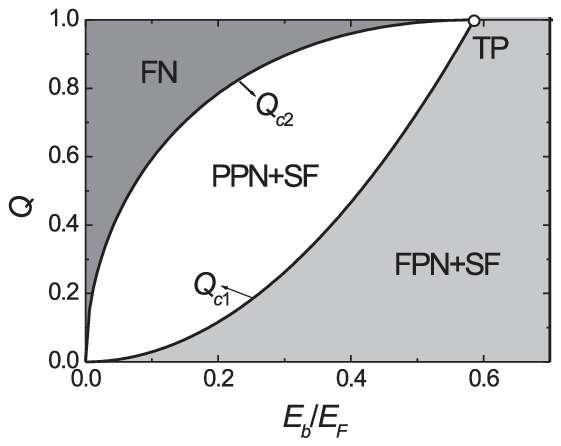,angle=0, height=160pt} \caption{Phase diagram for trapped 2D imbalanced fermion gases as a function of the population imbalance $Q$ (see Eq.~\eqref{polarization}) and $E_b/E_F$ with a \emph{zero} surface tension. At sufficiently low population imbalance the trap configuration (FPN+SF) consists of a SF core and a \emph{fully}-polarized normal shell. At intermediate population imbalance, the PPN+SF trap consists of a SF core and a \emph{partially}-polarized normal shell. Finally, in the FN configuration, the system consists of normal particles only. The critical phase boundaries $Q_{c1}$,and $Q_{c2}$ and the triple point TP are analytically described by
Eqs.~\eqref{tp}-\eqref{qc1}.\label{fig2}}
\end{figure}

We consider again trapped systems without surface tension ($\sigma=0$). In this case we are able to describe the entire phase diagram analytically. The phase diagram is mapped out in Fig.~\ref{fig2} as a function of the population imbalance $Q$ and the dimensionless parameter $E_b/E_F$. The diagram is characterized by the triple point TP which, as we show later, has the following coordinates:
\begin{align}
\label{tp}\left(  E_{b}/E_{F},Q\right)  _{TP}=\left(  2-\sqrt{2},\,1\right)  .
\end{align}

The triple point separates the FPN+SF, PPN+SF and FN trap configurations by
the lines $Q_{c1}$ and $Q_{c2}$. As follows from Eq.~\eqref{tp}, if the
interaction parameter is larger than $2-\sqrt{2}$, the trap always contains SF particles.

For values of $E_b$ below the TP and for small enough population imbalance $Q$, the FPN+SF configuration is the one of lowest energy, whereas for large enough $Q$, the trap configuration is always FN. At intermediate population imbalances and when the interaction strength is sufficiently weak ($E_b/E_F<2-\sqrt{2}$), a partially-polarized shell may be found in the N phase.
\begin{figure}
    \epsfig{figure=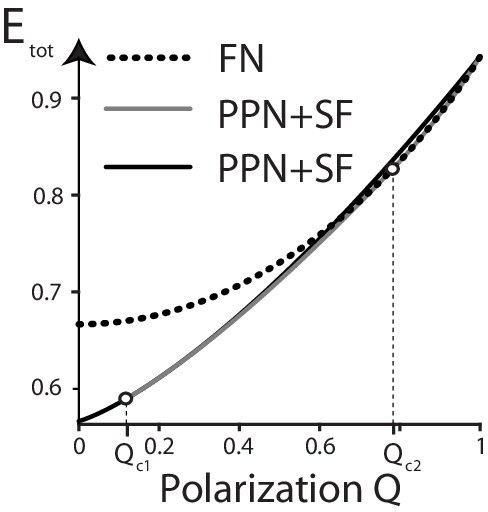,angle=0, height=170pt} \caption{Comparison of the total trap energy $E_{tot}$ of three relevant trap configurations against the population imbalance $Q$ in the case $E_b/E_F=0.2$ and with no surface tension. For low enough $Q<Q_{c1}=0.116$ the FPN+SF configuration has the lowest energy. For $Q_{c1}<Q<Q_{c2}=0.784$ it is the PPN+SF configuration that is stable, while a FN trap is encountered in the case $Q>Q_{c2}$. Both transitions are continuous.\label{fig3}}
\end{figure}
In Fig.~\ref{fig3} we illustrate the continuous nature of the transitions at $Q_{c1}$ and $Q_{c2}$ by plotting the total trap energy of the different trap configurations against the population imbalance $Q$ in case $E_b/E_F=0.2$. For population imbalances below $Q_{c1}=0.116$, the configuration of minimal energy is FPN+SF while for $Q_{c1}< Q<Q_{c2}$ it is the configuration with a partially-polarized normal shell (PPN+SF). Approaching $Q_{c2}$ from below, the equilibrium trap configuration is characterized by a vanishing population of SF particles.

For all lines in Fig.~\ref{fig2} we have obtained analytic expressions.
First of all, the critical line $Q_{c2}$ can be straightforwardly found since
we know that $\zeta$ vanishes at the transition. Putting $\zeta=0$ in
Eq.~\eqref{conditioning} yields~\cite{he}:
\begin{align}
\label{qc2}Q_{c2}=\left(  2-\frac{E_{b}}{E_{F}}\right)  \sqrt{\frac{E_{b}%
}{E_{F}}\left(  1-\frac{E_{b}}{4E_{F}}\right)  }.
\end{align}
The phase boundary between FPN+SF and N, on the other hand, is readily found
by equating Eqs.~\eqref{gieter} and~\eqref{deurmat}. It denotes the line
where, at the interface, the FNP becomes partially polarized and thus
$h=\left(  \mu_{SF}\right)  _{\zeta}$. As a result we arrive at~\cite{he}:
\begin{align}
\label{qc1}Q_{c1}=\left(  \frac{3+2\sqrt{2}}{2}\right)  \left(  \frac{E_{b}%
}{E_{F}}\right)  ^{2}.
\end{align}

\begin{figure}
    \epsfig{figure=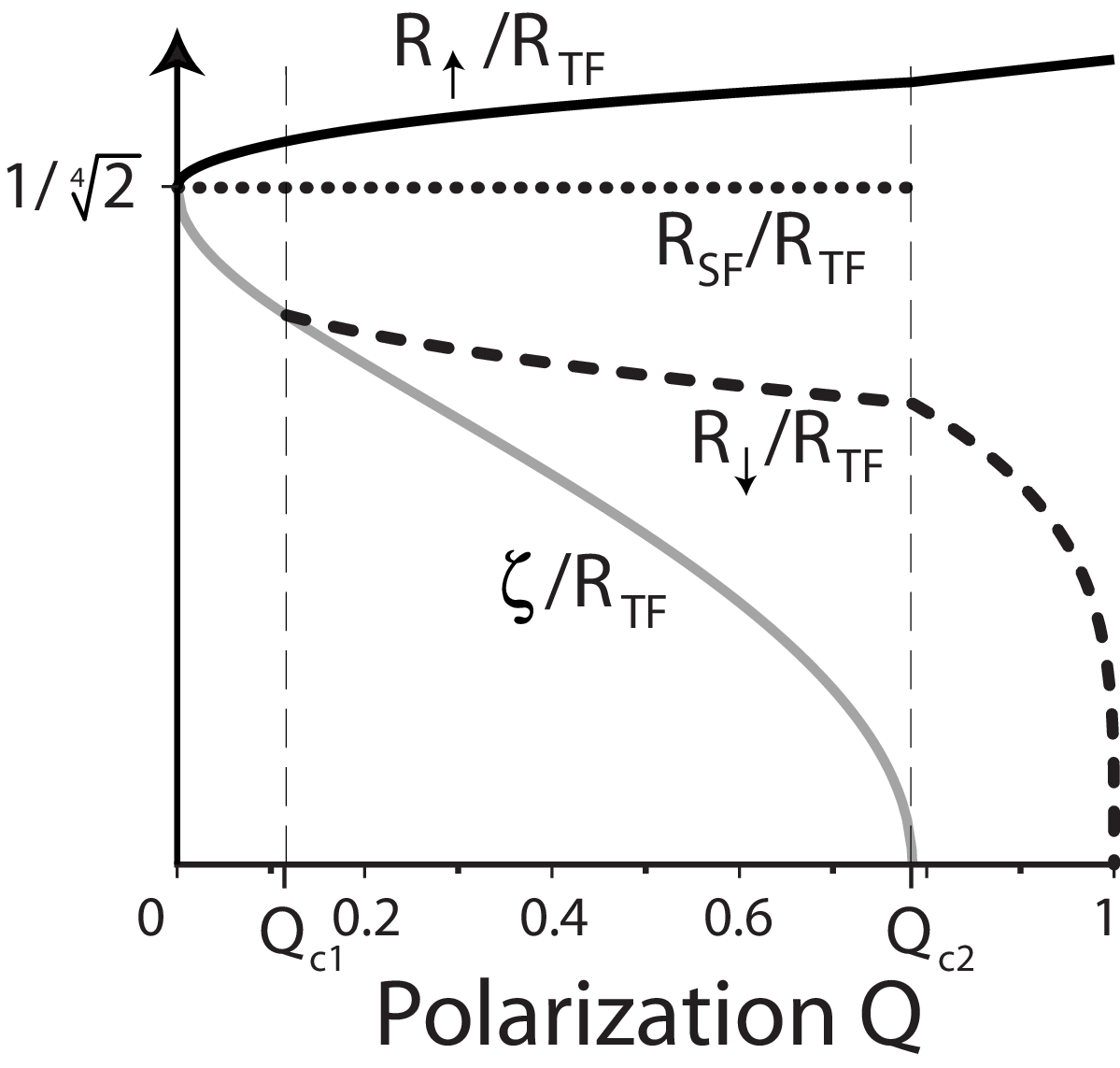,angle=0, height=170pt} \caption{The different radii $\zeta$, $R_{\downarrow}$, $R_{\uparrow}$ and $R_{SF}$ in units of $R_{TF}=\ell_{\omega}(8N)^{1/4}$ as a function of the population imbalance $Q$ in case $E_b/E_F=0.2$ and with a zero surface tension. \label{fig4}}
\end{figure}
Now, the triple point TP follow from Eqs.~\eqref{qc2} and~\eqref{qc1}. The phase diagram of Fig.~\ref{fig2} is in agreement with phase diagram of Fig.~4 in Ref.~\onlinecite{he} where it was derived assuming the transitions are critical.

The changes to the trap configurations at the transitions $Q_{c1}$ and $Q_{c2}$ are illustrated in Fig.~\ref{fig4}, where we plot the different radii as a function of the population imbalance $Q$ for $E_b/E_F=0.2$. For low enough population imbalance, $Q<Q_{c1}$, the interface of the FPN+SF configuration moves inwards with increasing population imbalance; the radii vary according to Eq.~\eqref{radiiPPN}. Note that the radius $R_{\downarrow}$ is not defined for the FPN+SF configurations as the N phase is fully polarized. On further increasing the population imbalance above $Q_{c1}$, the radius at which the N-SF interface lies, $\zeta$, decreases and finally vanishes at the transition point $Q_{c2}$.
\section{Collective Excitations} \label{collective} In this section we study the collective excitations of the imbalanced 2D fermion gases. The analogous excitations of a 3D system were already studied in the experiments of Ref.~\onlinecite{navon}. First, in Sect.~\ref{hydro} we take the approximation that both the N and the SF phases in the trap behave hydrodynamically. This allows us to treat a permeable interface, that is, an interface at which particles may go over from the N to the SF phase, as is present in case of a PPN+SF configuration. Then, in Sect.~\ref{collision}, we treat a collisionless N phase and are therefore able to treat exactly the collective excitations of the FN and, also of the FPN+SF configurations.

\subsection{Hydrodynamic Phases}\label{hydro} We begin by calculating the expressions for the bulk density fluctuations in the trap for both the SF and the N phase, assuming their dynamics can be described by hydrodynamics. Since we are interested in the lowest-energy frequencies of the trap when subjected to a weak excitation, we consider only the linearized hydrodynamic equations as usual.

Experimentally realized ultracold Fermi gases  are highly \emph{compressible} and, at each position, their densities deviate from their equilibrium value by an amount $\delta\rho_j$ with $j=\uparrow,\downarrow$ or $SF$. Henceforth, we will denote the deviations from equilibrium of the chemical potentials and the pressures by $\delta\mu$ and $\delta P$ respectively. An external perturbation of the system will induce the system to move at (position-dependent) velocity $\mathbf{v}$ in such a way as to satisfy the linearized Euler and the continuity equations~\cite{pethick,giorgini,heiselberg,bulgac,amoruso}:
\begin{subequations}
\begin{align}
        m\rho_j
        \partial_t\mathbf{v}_j&=-\boldsymbol{\nabla}\delta P_j- \delta\rho\boldsymbol{\nabla} V,\label{vijftiena}\\
        \partial_t\delta\rho_j&=-\boldsymbol{\nabla}\cdot(\rho_j\mathbf{v}),\label{vijftienb}
    \end{align}
\end{subequations}
where we only allow the particles to move in the $x-y$ plane such that we can consider $V(r)=m\omega_0^2r^2/2$. Combining Eqs.~\eqref{vijftiena} and~\eqref{vijftienb} and writing $\delta\rho_j(t)\rightarrow\delta\rho_j e^{-i\omega t}$, it follows that:
\begin{align*}
    m\omega^2\delta\rho_j=&-m\boldsymbol{\nabla}^2[c_j^2\delta\rho_j]-\boldsymbol{\nabla}\cdot(\delta\rho_j\boldsymbol{\nabla}V),
\end{align*}
where $c$ is the position-dependent velocity of sound which is introduced through the use of the Gibbs-Duhem relation~\cite{pethick,giorgini}. We continue by performing the transformation $\delta\rho_j(r,\theta)\rightarrow \delta\rho_j(r) e^{i\ell\theta}$ where $\ell$ is a positive integer or zero, so as to get:
\begin{align*}
    0=&\left[1-\widetilde{r}_j^2\right]\times\left[\frac{
    \partial^2}{
    \partial \widetilde{r}^2_j}+\frac{1}{\widetilde{r}_j}\frac{
    \partial}{
    \partial \widetilde{r}_j}-\frac{\ell^2}{\widetilde{r}_j^2} \right]\delta\rho_j\\
    &-2\widetilde{r}_j\frac{
    \partial \delta\rho_j}{
    \partial \widetilde{r}_j}+2\omega^2\delta\rho_j/\omega_0^2
\end{align*}
with $\widetilde{r}=r/R_j$. Solutions to this equation can be straightforwardly obtained in terms of the hypergeometric function $\text{F}$; moreover, they must be regular at $r=0$ when $j=SF$ and at $r=R_j$ when $j=\uparrow,\,\downarrow$. From Eq.~\eqref{chemisch}, it is clear that density deviations should be proportional to the chemical potential perturbations, or $\delta\rho\propto\delta\mu_j$. Thus
\begin{subequations}
    \begin{align}
        \delta\mu_{SF}&\propto r^\ell\text{F}\left(\alpha^+,\alpha^-,\alpha^0,(r/R_{SF})^2\right),\label{sfmu}\\
        \delta\mu_{i}&\propto r^\ell\text{F}\left(\alpha^+,\alpha^-,1,1-(r/R_{i})^2\right),
    \end{align}
\end{subequations}
with $i=\uparrow,\downarrow$, $\alpha^0=\ell+1$ and $2\alpha^\pm=\ell+1\pm[\ell^2 +2\omega^2/\omega_0^2]^{1/2}$.

As in Sect.~\ref{eos}, we must supplement these bulk equations by boundary equations at the N-SF interface. Mechanical equilibrium, as was expressed in Eq.~\eqref{laplace} for the equilibrium situation, is also demanded to be valid away from it. This results in the following expression:
\begin{align}
    \label{linlaplace} &\rho_{SF}\delta\mu_{SF}-\rho_{\uparrow}\delta\mu_{\uparrow} -\rho_{\downarrow}\delta\mu_{\downarrow}+\delta\zeta
    \partial_{r}\left(P_{SF}-P_{\uparrow}-P_{\downarrow}\right)\nonumber\\
    &=\sigma\delta\zeta\left(\frac{\ell^2-1}{\zeta^2}-\frac{3}{2\pi\rho_{SF}\ell_{\omega}^4}\right) +\frac{3m\sigma\delta\mu_{SF}}{2\pi\hslash^2\rho_{SF}\zeta}.
\end{align}
As mentioned before, particles may cross the permeable interface. At equilibrium, this was encoded by the chemical potential balance Eq.~\eqref{chemicalequilibrium}; away from equilibrium, this gives:
\begin{align}
    \label{pertchem} \delta\mu_{\uparrow}+\delta\mu_{\downarrow}-2\delta\mu_{SF} &=\delta\zeta\,
    \partial_{r}(2\mu_{SF}-\mu_{\uparrow}-\mu_{\downarrow}).
\end{align}
Lastly, the interconversion of particles between the N and the SF phase is not only restricted by chemical equilibrium (Eq.~\eqref{pertchem}), but also the continuity equation must be satisfied at the interface. Therefore, the fluctuating interface position $\delta\zeta$ depends on the velocities in the N and the SF as follows~\cite{andreev3}:
\begin{align}
    \label{parshin} \boldsymbol{e}_\zeta\cdot\left(2\mathbf{v}_{i}\rho_{i} -\mathbf{v}_{SF}\rho_{SF}\right)_{\zeta}=(2\rho_{i}-\rho_{SF})_\zeta
    \partial_t\delta\zeta,
\end{align}
with $i=\uparrow,\downarrow$ and $\boldsymbol{e}_\zeta$ is a unit vector perpendicular to the interface and directed towards the N side. Assume now that the N phase is fully polarized, such that $\left(\rho_{\downarrow}\right)_\zeta=0$. It is then easily shown that $\mathbf{v}_{SF}=\mathbf{v}_{\uparrow}=
\partial_t\delta\zeta$, implying that the interface is impermeable; the interface cannot move by interconversion of particles between the N and SF phase. In what follows we show that the collective excitations are strongly affected by the permeability of the interface. Taking the time derivative of Eq.~\eqref{parshin} and using the Euler equation yields:
\begin{align*}
    m\omega^2(2\rho_{i}-\rho_{SF})\,\delta\zeta=2\rho_{i}
    \partial_{r}\left(\delta\mu_{i}\right)-\rho_{SF}
    \partial_{r} \left(\delta\mu_{SF}\right).
\end{align*}

Prior to presenting our results, we outline the theory for treating a collisionless N phase.

\subsection{Collisionless Normal Phase}\label{collision} We briefly discuss how to treat the \emph{collisionless} N phase in a harmonic trap. We shall use the formalism presented in Refs.~\onlinecite{lazarides} and~\onlinecite{vanschaeybroeck3} and refer to these articles for details. In what follows, we treat the N phase as a non-interacting gas; this is valid when the N phase is fully polarized at the interface, that is, for all the FPN+SF configurations. As mentioned before, Eq.~\eqref{parshin} then implies that the interface is impermeable to particles.

The SF core, on the other hand, still behaves hydrodynamically, so that the chemical potential perturbation still satisfies Eq.~\eqref{sfmu}. Pauli blocking in the N gas, on the other hand, strongly suppresses interactions, leading to long local equilibration times and rendering a hydrodynamic description inapplicable. We must instead solve the Boltzmann-Vlasov equation for the full distribution function~\cite{lazarides}. The hydrodynamic description uses three numbers at each point, $\rho(\mathbf{r},t)$ and $\mathbf{v} (\mathbf{r},t)$ while the description of the N gas is based on the full distribution function $f(\mathbf{r},\mathbf{v},t)$; matching the two at the interface is nontrivial. The problem has been solved for the three-dimensional system in Ref.~\cite{lazarides}.

In brief, it is necessary to impose two boundary conditions: First, impermeability means that the interface velocity must equal the SF velocity:
\begin{align}
    \mathbf{e}_\zeta\cdot\mathbf{v}_{SF}=
    \partial_t\delta\zeta \quad\Rightarrow\quad
    \partial_{r} \left(\delta\mu_{SF}\right)=m\omega^2\delta\zeta.
\end{align}
Secondly, in the condition for mechanical equilibrium, Eq.~\eqref{linlaplace}, we must set $P_{\downarrow}=\rho_{\downarrow}=0$ and replace the pressure term $P_{\uparrow}$ by the momentum flux tensor $\delta \Pi_{rr}^\uparrow$. In 2D and for radially symmetric perturbations, a generalization of the calculations of Refs.~\onlinecite{lazarides} and~\onlinecite{vanschaeybroeck3} leads to:
\begin{align*}
    \delta \Pi_{rr\uparrow}&=\frac{\hslash\omega\delta\zeta (R_\uparrow^2-\zeta^2)^{3/2}}{\ell_{\omega}^3 \pi^2}\int_{0}^{1} d\chi\,\frac{\chi^3\cot(\omega\tau/2)}{\sqrt{1-\chi^2}},
\end{align*}
with
\begin{equation*}
    \tau=\arctan\left[2\chi\left(\sqrt{(R_\uparrow/\zeta)^2-1}-1/\sqrt{(R_\uparrow/\zeta)^2-1}\right)^{-1}\right]/\omega_0.
\end{equation*}
Solving these equations with respect to $\omega$ gives the frequencies of the system's \emph{breathing modes}.

\subsection{Results} 
\begin{figure}
    \epsfig{figure=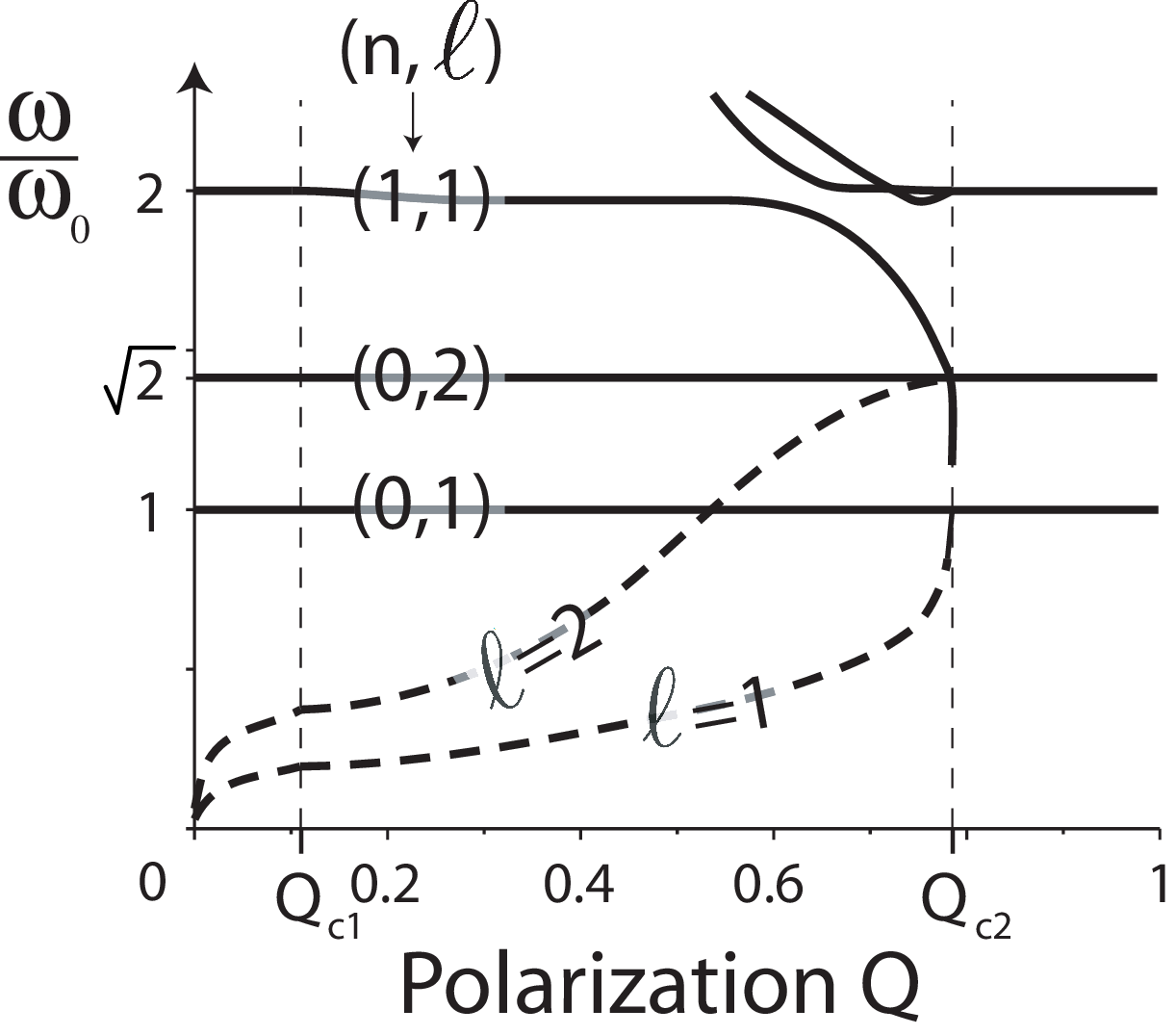,angle=0, height=190pt} \caption{The collective mode frequencies as a function of the population imbalance $Q$ in case the trap is fully hydrodynamic, $E_b/E_F=0.2$ and no surface tension is present. We show the in-phase (full lines) and out-of-phase (dashed lines) spectra of the $\ell=0,\, 1$ and $2$ modes. For low population imbalance $Q<Q_{c1}$ the system has a FPN+SF configuration while for $Q_{c1}<Q<Q_{c2}$, it is the PPN+SF configuration and it is FN for $Q>Q_{c2}$. A fully hydrodynamic formalism was used in order to be able to incorporate the permeable interface for the PPN+SF configuration.\label{fig6}}
\end{figure}

\begin{figure}
\epsfig{figure=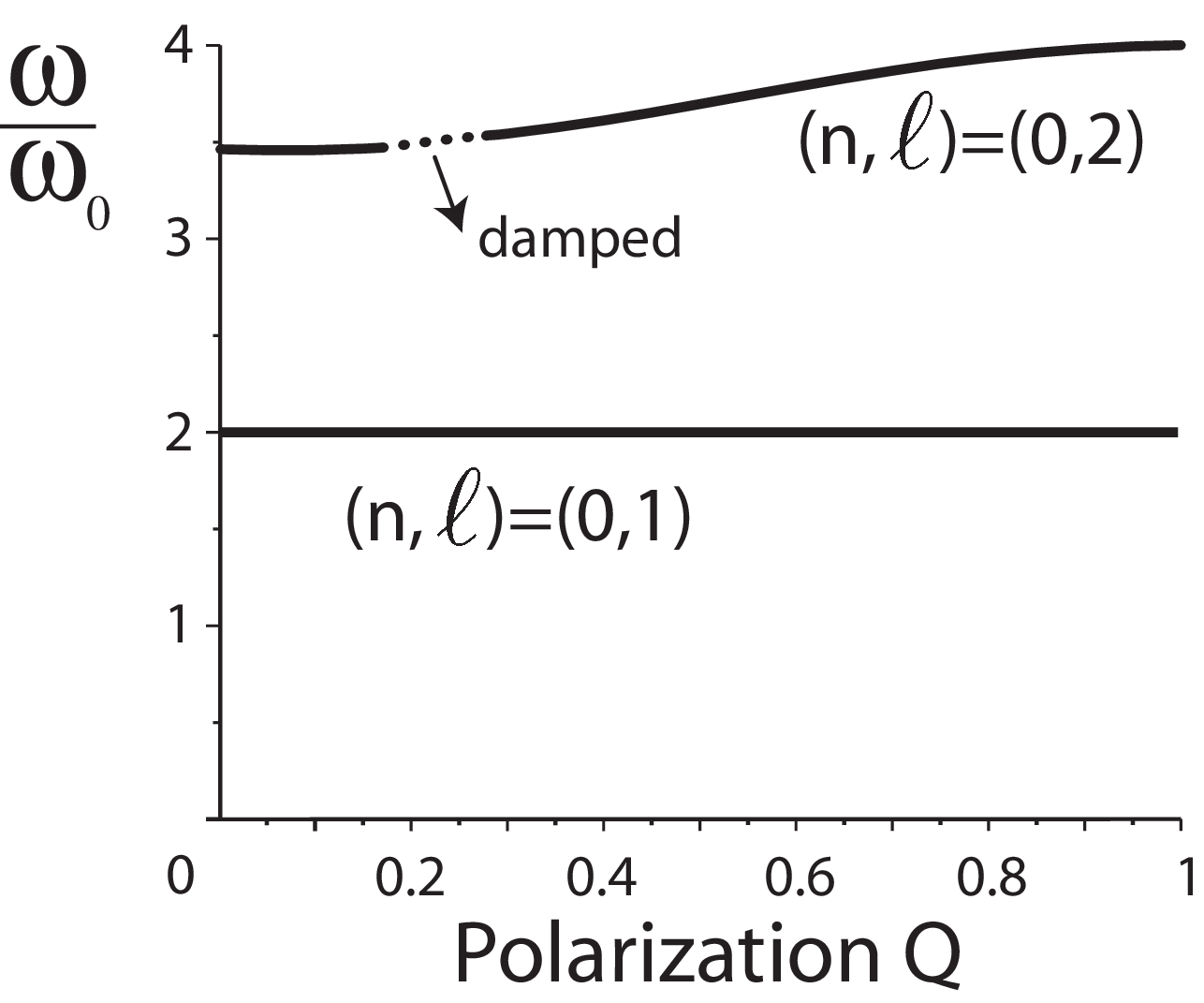,angle=0, height=160pt} \caption{The two lowest breathing mode frequencies as a function of the population imbalance $Q$ in case $E_b/E_F=0.6$ and with a collisionless normal phase and a zero surface tension. For the entire range of values of $Q$ the system has a FPN+SF configuration. The dashed region indicates collisionless interfacial damping.\label{fig7}}
\end{figure}

In Figs.~\ref{fig6} and~\ref{fig7} we present the spectra for the collective excitations of trapped imbalanced fermion gases in 2D taken into account a hydrodynamic (Sect.~\ref{hydro}) and collisionless N phase (Sect.~\ref{collision}), respectively.
We begin by discussing the fully hydrodynamic case of Fig.~\ref{fig6} which shows the results for a system with a zero surface tension ($\sigma=0$) and $E_b/E_F=0.2$. We distinguish two types of modes: the in-phase (IP) modes (full lines) and the out-of-phase (OOP) modes (dashed lines)~\cite{svidzinsky,lazarides}. The OOP modes are unique to a two-component system. In-phase and out-of-phase refer to the relative motion of the N-SF interface with respect to the outer boundary of the N shell (see Fig.~1 in Ref.~\onlinecite{lazarides}). Also, we can distinguish between three regimes depending on the trap configuration: from the phase diagram of Fig.~\ref{fig2}, it follows that for $Q<Q_{c1}=0.116$, the trap configuration is FPN+SF, while it is PPN+SF for $Q_{c1}<Q<Q_{c2}=0.784$ and FN for $Q>Q_{c2}$. At zero and full population imbalance, the single-species hydrodynamic spectrum $\omega^2=\omega_0^2[\ell+2n(n+\ell+1)]$ is recovered. For small but finite values of $Q$, OOP modes with low frequencies appear (note that we only show $\ell=1$ and $\ell=2$). Since the transitions at $Q_{c1}$ and $Q_{c2}$ are continuous, no discontinuities appear in the spectrum. For $Q>Q_{c2}$ the trap is fully N and the spectrum becomes independent of the imbalance. Finally, notice that the breathing mode (black line) which has a frequency of \emph{exactly} $\omega=2\omega_0$ when $Q<Q_{c1}$, becomes lower at $Q_{c1}$ and reappears for $Q>Q_{c2}$, that is when the SF core phase disappears.

In Fig.~\ref{fig7} we consider the lowest two \emph{breathing modes} of a system with a zero surface tension and with interaction parameter $E_b/E_F=0.5$; we treat the N particles collisionlessly. At zero polarization, we find the spectrum of the two lowest hydrodynamic breathing modes $2\omega_0$ and $2\sqrt{3}\omega_0$, while at full trap imbalance, we recover the collisionless breathing mode frequencies $2\omega_0$ and $4\omega_0$. As was also found in Fig.~\ref{fig6}, the lowest breathing mode of the FPN+SF configuration in Fig.~\ref{fig7} lies \emph{exactly} at $2\omega_0$ for all $Q$; we consider this feature to be an indication of the correctness of our formalism, even in the presence of a nonzero surface tension. For population imbalances $Q$ such that $0.116<Q<0.27$, the second mode is damped. This damping is caused by fermions which are resonantly driven at the interface as explained at length in Refs.~\onlinecite{lazarides} and~\onlinecite{vanschaeybroeck3}. 

\section{Energy Transport} 
We discuss now how heat is transferred throughout a trapped population-imbalanced fermion gas. Specifically, we focus on the heat transport through the N-SF interface~\cite{vanschaeybroeck4,parish}. Metastability due to a blocked transport of heat and spin was experimentally found to be the most important reason for explaining the discrepancies between early experiments on imbalanced fermion gases~\cite{hulet456,sommer}. We find that also in two dimensions, at low temperatures, energy flux is exponentially suppressed due to the presence of a SF gap. This is a consequence of the suppression of particles crossing the interface at low temperatures, since heat is transmitted through the N-SF interface when thermally excited particles penetrate the interface.

To understand this, consider a N particle incident on to the interface. If the N phase is partially polarized (for PPN+SF configurations), there are four scattering possibilities: Andreev reflection, specular reflection and hole-like and particle-like transmission~\cite{andreev}. A particle undergoing Andreev reflection pairs up with a particle of opposite spin to form a Cooper pair in the SF and therefore leaves behind a hole in the N phase. Particles penetrating the SF must have a minimum energy $\Delta_\zeta-h$, where $\Delta_\zeta$ is the SF gap at the interface~\cite{vanschaeybroeck4}. However, at sufficiently low temperature $T$, this penetration is rare because of the low statistical weight  $\mathrm{e}^{-(\Delta_\zeta-h)/k_BT}\ll 1$.

If, on the other hand, the N phase is fully polarized at the interface (for FPN+SF configurations), both Andreev reflection and quasiparticle transmission are suppressed so that all N particles  must be specularly reflected off the interface. Therefore no energy transport through the interface is possible for FPN+SF configurations~\footnote{This is valid only when the N phase is fully polarized at the interface. At nonzero temperature, however, the N phase may contain a nonzero fraction of $\downarrow$ particles. Nevertheless, the thermal conductivity remains very small due to the presence of the SF gap.}.

We now quanitfy this suppression of energy transport in the case of a PPN+SF configuration. To do this we calculate the heat resistivity of a N-SF interface on the basis of the Bogoliubov-de-Gennes equations, referring to Refs.~\onlinecite{vanschaeybroeck4} and~\onlinecite{vanschaeybroeck5} for details. A brief outline of the followed method is now given. We model the N-SF interface by its step-like behavior of the gap function: in the N phase $\Delta=0$, while according to mean-field theory and coexistence condition~\eqref{gieter}~\cite{tempere1}:
\begin{align*}
    \Delta_\zeta=\sqrt{2h^2+E_b^2/2}
\end{align*}
in the SF. Note that since we consider PPN+SF configurations, $h<E_b(1+\sqrt{2})/2$. Thermally excited particles and holes which are incoming on the N-SF interface are described by quasiparticle wave functions. Matching the wave functions and their derivatives at the interface allows us to calculate the transmission coefficients. We then calculate the heat conductivity $\kappa$ of the N-SF interface which is the net heat flux per temperature difference across the interface. We find that in case a small temperature bias between N and SF phase exists, the thermal equilibration process will be blocked by a thermal conductivity that decays exponentially fast as the temperature falls. In the limit of very weak interactions $E_b/\mu_{SF}\ll 1$ and at low temperatures $k_BT\ll\Delta_\zeta-h$, one can use the Andreev approximation to obtain the following analytic expression for the two-dimensional case~\cite{andreev,andreev2,mcmillan}:
\begin{align}\label{andrapp}
\frac{\kappa}{\kappa_{N}}\approx\sum_{\sigma=\pm}\frac{e^{-(\Delta_\zeta-\sigma h)/k_BT}\sqrt{\pi}(\Delta_\zeta-\sigma h)^2}{k_BT\sqrt{2k_BT\Delta_\zeta}}.
\end{align}
It is clear that the dominant factor at small temperatures is $e^{-(\Delta_\zeta-h)/k_BT}$. We have found that the approximation of Eq.~\eqref{andrapp} is in very good agreement with numerically obtained results, even for large values of $E_b/\mu_{SF}$.

We therefore conclude that, at sufficiently low temperatures, thermal equilibration between N and SF phases at different temperatures will be considerably slowed down due to the SF gap.

\section{Discussion and Conclusion}

Although our formalism is only quantitatively correct in the BEC regime, we expect that it also yields qualitatively realistic information throughout the entire BEC-BCS crossover~\cite{zhang1}. The condition $E_b\ll \mu_{SF}$ is generally correct in trap center.

We have studied population-imbalanced Fermi mixtures under quasi two-dimensional confinement. Using results previously published for the homogeneous case and using the local-density approximation, we have shown that there are three relevant configurations of the trapped cloud: fully normal (FN), a partially-polarized normal cloud surrounding a superfluid core (PPN+SF) and a fully polarized normal cloud surrounding a superfluid core (FPN+SF).

We first show that, in the absence of surface tension, the system is completely specified by two numbers: the polarization, $Q=(N_{\uparrow}-N_{\downarrow})/(N_{\uparrow}+N_{\downarrow})$ (where $N_i$ the total number of particles of species $i$), and the interaction parameter $E_b/E_F$ (see Eqs.~\eqref{boundenergy} and~\eqref{Fermi} for definitions). Thus, the phase diagram we obtain (Fig.~\ref{fig2}) is \emph{universal}, in the sense that it only depends on a small number of parameters. This phase diagram has the property that, for $E_b/E_F>2-\sqrt{2}$ (see Eq.~\eqref{tp}) there is no critical $Q_c$ above which the SF disappears; in other words, above this $E_b/E_F$, there is SF in the trap for all polarizations.

We then turn to the calculation of collective excitation modes of the system. We study these in two approximations.

First we take both the SF and the N parts to behave hydrodynamically. In the FPN+SF and the PPN+SF case we find out-of-phase modes, typical of two-component systems.

We then study the more physically relevant case where the fully polarized N gas displays collisionless behaviour. In this approximation we calculate the collective mode frequencies and find the expected limits, as well as damping related to Landau damping for a range of polarizations.

Note that the finite-temperature corrections to the collective excitation frequencies obtained here can be found in Ref.~\cite{klimin44}.

Finally we discuss the effect of the interface on thermal transport through the cloud. This would be important in studying how the trapped cloud equilibrates starting from an initially nonuniform temperature distribution, such as might be obtained while the system is being cooled. Based on previously published work, we find that thermal transport is strongly suppressed at low temperatures due to the presence of the gap in the excitation spectrum inside the SF.

\section{Acknowledgements} We acknowledge partial support by Project No.~FWO G.0115.06. B.V.S. and A.L. acknowledge support from FWO and the Netherlands Foundation for Scientific Research (NWO), respectively. We would like to thank Joseph Indekeu for useful discussions.

\end{document}